\documentclass[twoside,fleqn]{article}
\usepackage{espcrc2}
\usepackage{amssymb,amsmath}
\usepackage{graphicx}

\newcommand{\bbra}{\langle\!\langle}
\newcommand{\kket}{\rangle\!\rangle}

\newcommand{\psibar}{\bar{\psi}}
\newcommand{\bQ}{\bar{Q}}

\newlength{\colw}
\begin{document}
\makeatletter \@mathmargin = 0pt \makeatother
\bibliographystyle{h-elsevier2}

\title{Improving Algorithms to Compute All Elements of the
Lattice Quark Propagator}

\author{Alan \'O Cais\address[tcd]{School of Mathematics, Trinity
College, Dublin 2, Ireland}\thanks{This contribution is based on
parallel talks given by Alan \'O Cais and K.~Jimmy Juge at Lattice
2004.}, 
K.~Jimmy Juge\addressmark[tcd]$^*$, 
Mike J.~Peardon\addressmark[tcd], 
Sin\'ead M.~Ryan\addressmark[tcd], 
Jon-Ivar Skullerud\addressmark[tcd] }

\begin{abstract}
We present a new exact algorithm for estimating all elements of the
quark propagator. The advantage of the method is that the exact
all-to-all propagator is reproduced in a large but finite number of
inversions. The efficacy of the algorithm is tested in Monte Carlo
simulations of Wilson quarks in quenched QCD. Applications that are
difficult to probe with point propagators are discussed.
\end{abstract}

\maketitle

\section{INTRODUCTION}
The discovery of exactly chirally symmetric fermions on the lattice
has triggered intensive research in the development of algorithms to
simulate dynamical Ginsparg-Wilson fermions on the lattice. In light
of recent work in this area, we anticipate that there will be a
limited number of expensive, full QCD configurations in the near
future. One would like to extract all the information that one can
from these lattices without being restricted by point propagators,
which would appear highly wasteful considering the cost to generate the
configurations.  Point propagators do not require massive computing
power but restrict the physics one has access to, mainly
the flavour non-singlet spectrum. They also restrict the interpolating
operator basis used to produce early plateaux in effective
masses, for instance, since a new inversion must be performed for
every operator that is not restricted to a single lattice
point. Variational methods would be much more powerful
with the use of all-to-all propagators.

All-to-all propagators \cite{Kuramashi:1993ka,Dong:1993pk,deDivitiis:1996qx,Michael:1998sg,Duncan:2001ta,Peardon:2002ye} provide a solution to these problems, but are
usually too expensive to compute exactly as this requires an unrealistic
number of quark inversions. Stochastic estimates tend to be very noisy
and variance reduction techniques are crucial in order to separate the
signal from the noise.  In this paper we propose an exact algorithm to
compute the all-to-all propagator utilizing the idea of low-mode
dominance corrected by a stochastic estimator which yields the exact
all-to-all propagator in a finite number of quark inversions.

\section{THEORY AND NOTATION}
\subsection{Spectral Decomposition}
Theoretical arguments, backed up by numerical evidence
\cite{Duncan:1998gq,Neff:2001zr}, indicate
that the low lying eigenmodes of the Hermitian Dirac operator captures
much of the important infrared physics in hadronic interactions. It
would be desirable to take advantage of this fact and solve exactly
for as many of the eigenmodes as possible to estimate the all-to-all
quark propagator.

To this end, we define the Hermitian Dirac operator $Q=\gamma_5M$
where $M$ is the usual Dirac operator. The truncated representation of
the quark propagator is then given by,
\begin{equation}\label{eq:trunc}
\bQ_0=\sum_{i}^{N_{ev}}\frac{1}{\lambda_i}v^{(i)}(\vec{y},y_0)\otimes
 v^{(i)}(\vec{x},x_0)^\dag\,,
\end{equation}
where $Qv^{(i)}=\lambda_iv^{(i)}$. The truncation at an arbitrary
number, $N_{ev}$, of eigenvectors, however, leaves the theory
non-unitary, making it mandatory to correct it. 

\subsection{Dilution (Noisy Estimators)}
The standard method of estimating the all-to-all quark propagator is
to sample the vector space stochastically. One generates an ensemble
of random noise vectors, $\{\eta_{[1]},\cdots,\eta_{[N_r]}\}$, with
the property
\begin{equation}\label{eq:mixing}
\bbra\eta(x)\!\otimes\!\eta(y)^\dag\kket\propto\delta_{x,y}\,,
\end{equation}
where $\bbra\cdots\kket$ denotes average over noise samples.
The solution for each vector is obtained in the usual way,
\begin{equation}
\psi_{[r]}(x)=Q^{-1}\eta_{[r]}(y)\,.
\end{equation}
The all-to-all quark propagator is estimated as 
\begin{equation}
\begin{split}
Q^{-1}
 &=\langle\!\langle\psi\!\otimes\!\eta^\dagger\rangle\!\rangle \\
 &\simeq\frac{1}{N_r}\sum_r^{N_r}\psi_{[r]}(y)\otimes\eta_{[r]}(x)^\dag\,.
\end{split}
\label{eq:noisy}
\end{equation}
This method is noisy because it relies on delicate cancellations in
the ${\mathcal O}(1)$ noise over many
samples to find the signal, which falls off exponentially with the
separation. We propose to remove the ${\mathcal O}(1)$ random noise by
``diluting'' the noise vector according to some dilution scheme
resulting in an exponential gain in the variance. A particularly
important example of dilution for measuring temporal correlations in
hadronic quantities is ``time dilution" where the noise vector is
broken up into pieces which only have support on a single timeslice
each,
\begin{equation}
\eta(\vec{x},t)=\sum_{j=0}^{Nt-1}\eta^{(j)}(\vec{x},t) \, ,
\end{equation}
where $\eta^{(j)}(\vec{x},t)=0$ unless $t=j$.

\begin{figure}
\includegraphics[width=\colw]{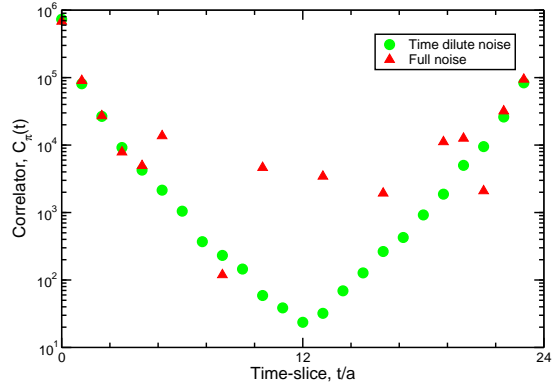}
\vspace{-6mm}
\caption{The psuedoscalar propagator computed with and without time dilution.}
\label{fig:timedil}
\end{figure}

\begin{figure}
\includegraphics[width=\colw]{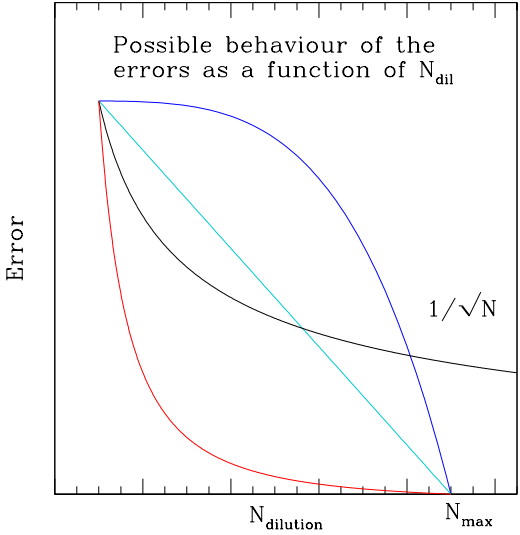}
\vspace{-6mm}
\caption{A cartoon of possible deviations of the stochastic estimates
  of the exact solution (at $N_d=N_{\text{max}}$) for different dilution
  paths.  Simply adding noise vectors will give a $1/\sqrt{N}$ behaviour.}
\label{fig:cartoon}
\end{figure}

Each diluted source is inverted resulting in $N_t$ pairs of vectors,
$\{\psi^{(i)}(\vec{x},t),\eta^{(i)}(\vec{x},t)\}$, which then gives an
unbiased estimator of $\langle\psi\psibar\rangle$ with a single noise
source,
\begin{equation}
Q^{-1}=\sum_{i=0}^{N_t-1}\psi^{(i)}(\vec{x},t)\otimes\eta^{(i)}(\vec{x}_0,t_0)^\dag\,.
\end{equation}
We show the effect of time dilution on a pseudoscalar propagator on a
$12^3\times24$ lattice in Fig.~\ref{fig:timedil}. The triangles are
the average of 24 noise sources without any dilution and the circles
are from a single noise source which has been time-diluted.  This
particular scenario is analogous to the ``wall source on every
timeslice" method used by the authors of Ref.~\cite{Kuramashi:1993ka}
to estimate the disconnected diagrams appearing in hadronic scattering
length calculations. Our method is, however, more general and can be
extended to the spin, color and space components of the source
vector. The ``homeopathic'' limit of this dilution procedure results
in the {\it exact} all-to-all propagator in a finite number of steps
(Fig.~\ref{fig:cartoon}). This limit cannot be reached in practice on
realistic lattices, but the path of dilution may be optimized so that
the noise from the gauge fields dominate the errors in the hadronic
quantities of interest with only a small, managable number of fermion
matrix inversions.

\subsection{`Hybrid' Method}
The evidence that much of the hadronic physics we would like to
extract from the lattice is in the low-lying eigenmodes of the
(Hermitian) Dirac operator suggests that one try to calculate as many
as possible of the low modes exactly and correct for the truncation
with the noisy method. This gives rise to two concerns: firstly, the
correction should leave the exactly solved low-lying modes intact; and
secondly, it should not introduce large uncertanties in the
process. We propose that the stochastic method with noise dilution is
a natural way to accomodate both of those concerns.

First, we note that the exact $N_{ev}$ low eigenmodes obtained
separately naturally divide $Q$ into two subspaces, $Q=Q_0+Q_1$,
defined by
\begin{align}
Q_0&=\sum_{i=1}^{N_{ev}}\lambda_iv^{(i)}\otimes v^{(i)\dagger}\,, \\
Q_1&=\sum_{j=N_{ev}+1}^{N}\lambda_jv^{(j)}\otimes v^{(j)\dagger}\,.
\end{align}
Similarly, the quark propagator is broken up into two pieces,
$Q^{-1}=\bQ_0+\bQ_1$,
where $\bQ_0$ is given by Eq.~(\ref{eq:trunc}) and $\bQ_1$ is the
remaining unknown piece. We correct for the truncation and estimate
$\bQ_1$ using the stochastic method,
$\bQ_1=\langle\!\langle\psi\!\otimes\!\eta^\dagger\rangle\!\rangle$
with $N_r$ noise vectors, $\{\eta_{[1]},\cdots,\eta_{[N_r]}\}$. The
solutions are given by
\begin{equation}\label{eq:remain}
\psi_{[r]}=\left(Q^{-1}{\mathcal P}_1\right)\eta_{[r]}=Q^{-1}\left({\mathcal P}_1\eta_{[r]}\right)\,,
\end{equation}
where ${\mathcal P}_1$ is the projection operator
\begin{equation}
{\mathcal P}_1={\bf 1}-{\mathcal P}_0={\bf 1}-\sum_{j=1}^{N_{ev}}v^{(j)}\otimes v^{(j)\dagger}\,.
\end{equation}
Note that Eq.~(\ref{eq:remain}) follows from the identity $Q^{-1}{\mathcal P}_1=\bQ_1$. 
As was mentioned earlier, the idea of dilution will be applied to the
stochastic estimation of $\bQ_1$. Each random noise vector,
$\eta_{[r]}$, that is generated will be diluted and orthogonalized
(wrt ${v^{(i)}}$) so that it can be used to obtain $\psi_{[r]}$. In
other words, we now have the following set of noise vectors:
\begin{equation}
\left\{\left(\eta_{[1]}^{(1)},\cdots,\eta_{[N_r]}^{(1)}\right),\cdots,\left(\eta_{[1]}^{(N_d)},\cdots,\eta_{[N_r]}^{(N_d)}\right)\right\}\,,\notag
\end{equation}
where the upper indices denote the dilution and the lower indices
label the different noise samples. We note that the noise vectors are
mutually orthogonal to each other due to the dilution before an average
over different random vectors are taken, i.e.,
\begin{equation}
\eta_{[r]}^{(i)}(\vec{x},t)\otimes\eta_{[s]}^{(j)}(\vec{y},t^\prime)^\dagger\propto\delta^{ij}.
\end{equation}
This results in smaller variance than the standard method which mixes
noise, as Eq.~(\ref{eq:mixing}) shows.

\begin{figure}
\includegraphics[width=\colw]{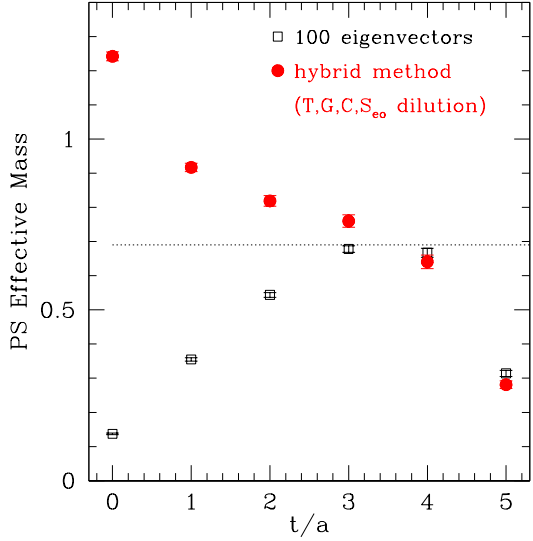}
\vspace{-6mm}
\caption{The effective mass from 100 eigenvectors and from the hybrid
  method with 100 eigenvectors and a time-diluted noise vector. The lattice size is $6^3\times12$.}
\label{fig:hybrid}
\end{figure}

There is a natural way to combine the two methods to estimate the
all-to-all quark propagator. The similarity in the structure of
Eq.~(\ref{eq:trunc}) and Eq.~(\ref{eq:noisy}) suggests that one
construct the following ``hybrid list'' for the source and solution
vectors:
\begin{align}
w^{(i)}&=\biggl\{\frac{v^{(1)}}{\lambda_1},\cdots,\frac{v^{(N_{ev})}}{\lambda_{N_{ev}}},\eta^{(1)},\cdots,\eta^{(N_d)}\biggr\}\\
u^{(i)}&=\biggl\{v^{(1)},\cdots,v^{(N_{ev})},\psi^{(1)},\cdots,\psi^{(N_d)}\biggr\}
\end{align}
where the indices run over $N_{HL}=N_{ev}+N_d$ elements.
The master formula for the unbiased, variance reduced estimate of the
all-to-all quark propagator is then given by
\begin{equation}
M^{-1}=\sum_{i=1}^{N_{HL}}u^{(i)}(\vec{x},x_4)\otimes w^{(i)}(\vec{y},y_4)^\dagger\gamma_5\,.
\end{equation}
Using the pion as an example, we can demonstrate how unitarity is
recovered from the truncated propagator. In Fig.~\ref{fig:hybrid}, we
show the effective mass from the truncated propagator and from the
hybrid method with a time, spin, color and space (even-odd) diluted 
noise vector. The truncated propagator, which approaches the 
asymptotic value from below, is corrected by the addition of the 
diluted noisy propagator. 

\section{QCD IMPLEMENTATION}
All-to-all propagators not only expand the range of applications
accessible in lattice QCD but also considerably simplify the
construction of hadronic interpolating operators. A pseudoscalar
correlator is constructed in the following way with
traditional, point-to-all quark propagators,
\begin{equation}
C=M^{-1}(\vec{x},t;\vec{0},0)^{\dagger\ ij}_{\ \alpha\beta}
 \;M^{-1}(\vec{x},t;\vec{0},0)^{ji}_{\beta\alpha}\,.
\end{equation}
The construction of a better operator is awkward at best since the
operator at timeslice $t$ is constructed from various components of
the quark propagator connecting the source and sink. All-to-all
propagators eliminate this complication as the operator at timeslice
$t$ is constructed from vectors on that timeslice,
\begin{equation}
\begin{split}
{\mathcal O}_{PS}^{(i,j)}(\vec{x},t)
 &=(w^{(i)}_{[r_1]}(\vec{x},t)^\dagger\gamma_5)\gamma_5u^{(j)}_{[r_2]}(\vec{x},t)\\
 &=w^{(i)}_{[r_1]}(\vec{x},t)^\dagger u^{(j)}_{[r_2]}(\vec{x},t)\,,
\end{split}
\end{equation}
where the extra factor of $\gamma_5$ comes from the use of the
Hermitian Dirac matrix, $\gamma_5M$.
For complicated operators such as those used to project out
hybrid/exotic states, this is a much needed simplification. For
example, an interpolating operator for the exotic hybrid $1^{-+}$ can
be constructed from combinations of gluonic paths projecting out the
relevant quantum numbers \cite{Lacock:1996vy}. One term in such a sum
may be
\begin{equation}
w^{[i]}(\vec{x})^\dag U_z(\vec{x})U_y(\vec{x}+\hat{z})U_z^\dagger(\vec{x}+\hat{e}_y) u^{[j]}(\vec{x}+\hat{e}_y)\,,
\end{equation}
where all the variables are located on a single timeslice.
A standard P-wave state may for example be constructed as follows,
\begin{equation}
{\mathcal O}_{P}^{(i,j)}(\vec{x},t)=w^{(i)}(\vec{x},t)^\dagger(D_ku^{(j)})(\vec{x},t)\,,
\end{equation}
where $D_k$ is the covariant derivative.

Using all-to-all propagators, correlation functions are also
constructed in an intuitive manner. Hadronic correlation functions
are obtained by correlating interpolating operators
sitting at different timeslices, e.g.
\begin{equation}
C(t,t_0)=\sum_{i,j}^{N_{HL}}{\mathcal O}_{[r_1,r_2]}^{(i,j)}(t){\mathcal O}_{[r_2,r_1]}^{(j,i)}(t_0)\,
\end{equation}
for isovector two-point correlators (propagators).

Noting that this is simply a contraction of the source and solution
vectors with an outer sum over the hybrid list indices, this gives
even greater scope for simplification. Once the machinery of the
hybrid list summation is set up in the program, it becomes a black box
to the end user who simply supplies the subroutines to create the
required operators from the quark, antiquark and gluon fields on a
timeslice.

For isoscalar mesons, the disconnected part of the propagator must be
included, yielding the following contraction,
\begin{multline}
\left\{w^{(i)}_{[2]}(t)^\dag\gamma_5\Gamma u^{(j)}_{[1]}(t)\right\}
\left\{w_{[1]}^{(j)}(t_0)^\dag\gamma_5\Gamma^\dag u_{[2]}^{(i)}(t_0)\right\}\\
-\left\{w_{[1]}^{(j)}(t)^\dag\gamma_5\Gamma u_{[1]}^{(j)}(t)\right\}
\left\{w_{[2]}^{(i)}(t_0)^\dag\gamma_5\Gamma^\dag u_{[2]}^{(i)}(t_0)\right\}\,.
\notag
\end{multline}

\section{TESTS}
\begin{figure}
\includegraphics[width=\colw]{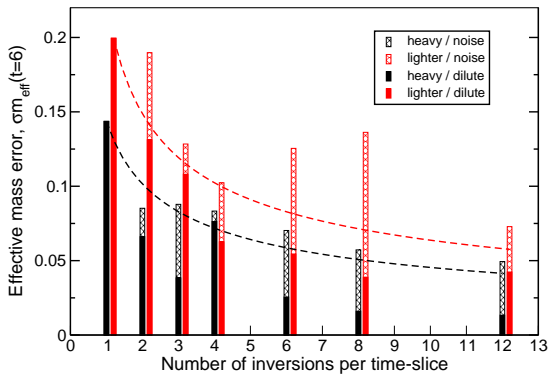}
\vspace{-6mm}
\caption{Equal cost test of the pion effective mass at timeslice 6.}
\label{fig:equalcost}
\end{figure}

We use the Wilson action to illustrate the effectiveness of
the variance reduction although it is expected to work even better for
a chiral action and light quarks. The simulation parameters are shown in
Table~\ref{table:kappas}.
\begin{table}
\caption{Fermion run parameters.  The values for $m_\pi, m_\rho$ are
  taken from Ref.~\protect\cite{Butler:1994em}.}
\label{table:kappas}
\begin{tabular}{rrrrr}
\hline
 $n_x^3\times n_t$ & $\kappa$ & $am_\pi$ & $am_\rho$ & $m_\pi/m_\rho$ \\ \hline
$ 6^3\times12$ & 0.1600 & 0.69 & 0.80 & 0.86 \\
$12^3\times24$ & 0.1600 & 0.69 & 0.80 & 0.86 \\
$12^3\times24$ & 0.1663 & 0.38 & 0.62 & 0.61 \\
$12^3\times24$ & 0.1675 & 0.30 & 0.60 & 0.50 \\\hline
\end{tabular}
\end{table}
In order to demonstrate the efficacy of the noise dilution method, we
perform an ``equal cost'' comparison test. Equal cost here shall mean
the same number of inversions performed, disregarding the sparseness
of the source vector. For example, an even-odd dilution in space
results in twice as many inversions as the one without, hence two
independent sets of time diluted noise vectors are generated to
compare the noise in the effective mass of the psuedoscalar.  As we
have seen that no signal is obtained otherwise, all noise vectors have
been time diluted. The results are shown in
Fig.~\ref{fig:equalcost}. The noise diluted errors consistently show
smaller errors than the standard stochastic method.
At this point, one may note that the isovector correlation function
(when both quarks are saved as all-to-all) involves a large number
($N_{HL}\times N_{HL}$) of contractions. We can take advantage of
having saved different random source/solution samples by reusing them
in the contraction. In other words, one can generate $N_R$ samples of
noise vectors, $\eta_{[r]}$, and save the corresponding solutions,
$\psi_{[r]}$ to disk and perform the contraction,
\begin{equation}
C=\sum_{r,s}w^\dagger_{[r]}u_{[s]}(t)w^\dagger_{[s]}u_{[r]}(t_0)\,,
\end{equation}
yielding $\sim\!N_R^2$ samples of the pion correlation
function. The errors correspondingly decrease faster than the na\"ive
$1/\sqrt{N_R}$, although the measurements are correlated. We
have seen in our preliminary tests that the gain in error reduction is
comparable to some dilution choices. It is clear that if one can
afford to save the noise/solution vectors onto disk, then this is a
straightforward method of variance reduction.
 
\begin{figure}[t]
\includegraphics[width=\colw]{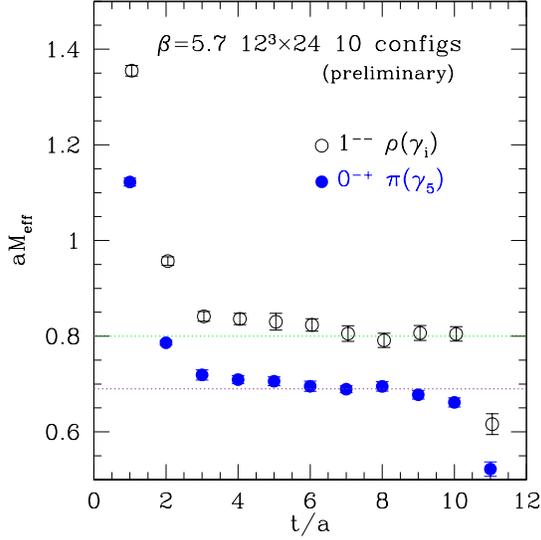}
\vspace{-7mm}
\caption{The pion and rho meson effective masses from 10
  configurations on a $\beta=5.7$, $12^3\times24$ lattice.}
\label{fig:spectro1}
\end{figure}

\begin{figure}[t]
\includegraphics[width=\colw]{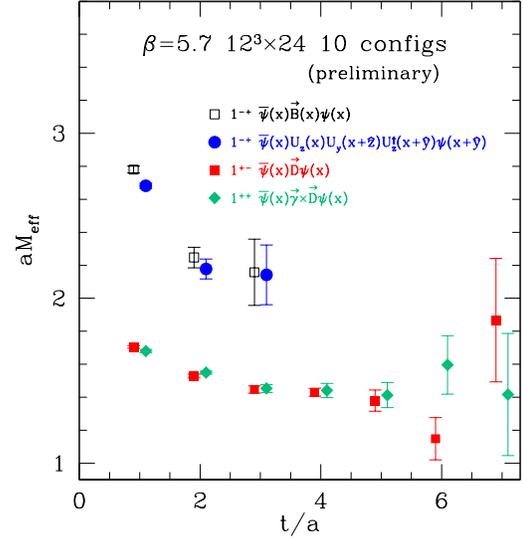}
\vspace{-6mm}
\caption{Preliminary results for a variety of P-wave states. The
  staple hybrid operators actually use a length-2 staple.}
\label{fig:spectro2}
\end{figure}

We have performed some spectroscopy calculations using the idea of
dilution.  Standard meson spectroscopy including pions, rhos, heavy
exotic hybrids (with two different operators) and standard P-wave
states were studied. For the P-wave states in particular, it is  difficult to get a good overlap using point propagators.  The effective masses for
all the particles are shown in Fig.~\ref{fig:spectro1} and
Fig.~\ref{fig:spectro2}.  Although in this test study we only used 10
quenched gauge configurations, we have a surprisingly good signal for
P-waves as well as the extremely heavy hybrid states.  Furthermore,
decay constants can be extracted from point--smeared and
smeared--smeared correlators if all-to-all propagators are used
\cite{Ryan:2004xx}.

\section{SUMMARY}
We have presented a new algorithm to estimate the all-to-all
propagator.  All-to-all propagators make it possible to make use of
all the available information in a gauge configuration, which
considering the cost to generate full QCD configurations may be of
crucial importance. They are also a necessary ingredient in flavour
singlet physics.  

All-to-all propagators have a further advantage over point propagators
in that operator construction is considerably simplified: The
operators are constructed in a natural way from local fields, and
extended operators used in variational methods may be employed at no
additional cost.

We have presented evidence that diluting stochastic estimators in
time, colour, spin or other variables gives less variance than
traditional noisy estimators.  This is not unexpected, since dilution
will yield the exact all-to-all propagator in a finite number of
iterations.  More work is needed to determine the optimal dilution
path for different observables.

The hybrid method allows one to extract the important physics from
low-lying eigenmodes and combine this with a noisy correction in a
natural way.  This may be implemented in such a way that the user can
be blind to the details of the dilution and eigenvector list.

Preliminary tests of the algorithm have been presented with small
lattices and a small number of configurations. We find that the
algorithm works as expected for standard light-light, static-light and
exotic meson spectroscopy. Further work will include higher statstics on
lighter masses as well as tests on disconnected diagrams.

\section*{ACKNOWLEDGEMENTS}
This work was funded by an IRCSET award SC/03/393Y and the IITAC
PRTLI initiative.
\bibliography{trinlat_bib/trinlat}

\end{document}